\documentclass[aps,prl,preprint,12pt]{elsarticle}
\usepackage{graphicx}
\usepackage[margin=1.0in]{geometry}
\usepackage{color, colortbl}
\usepackage{hyperref}
\usepackage{float}

\title{CLAS12 Track Reconstruction with Artificial Intelligence}

\author[1]{Gagik Gavalian}
\author[2]{Polykarpos Thomadakis}
\author[2]{Angelos Angelopoulos}
\author[2]{Nikos Chrisochoides}
\address[1]{Jefferson Lab, Newport News, VA, USA}
\address[2]{CRTC, Department of Computer Science, Old Dominion University, Norfolk, VA, USA}

\begin{document}
\begin{abstract}
  In this article, we describe the implementation of Artificial Intelligence models in track reconstruction
  software for the CLAS12 detector at Jefferson Lab. The Artificial Intelligence-based approach resulted 
  in improved track reconstruction efficiency in high luminosity experimental conditions.  The track
 reconstruction efficiency increased by $10-12\%$ for a single particle, and statistics in multi-particle physics 
 reactions increased by $15\%-35\%$ depending on the number of particles in the reaction. The implementation 
 of artificial intelligence in the workflow also resulted in a speedup of the tracking by $35\%$. 

\end{abstract}

\maketitle
%\end{CJK*}

\section{Introduction}
\indent

Nuclear Physics experiments have become increasingly complex over the past decades, with more complex detector systems and higher luminosities. In emerging experiments where detector occupancies are higher, there is a need for new approaches to data processing that can improve data reconstruction accuracy and speed. New developments in the Artificial Intelligence (AI) field present promising alternatives to conventional algorithms for data processing. Machine Learning (ML) algorithms are being employed in various stages of experimental data processing, such as detector data reconstruction, particle identification, detector simulations, and physics analysis. 

In this paper, we present the implementation of machine learning models into the CLAS12 charged-particle track reconstruction software. Detailed analysis 
of the reconstruction, performance is presented, comparing track reconstruction efficiency and speed improvements to conventional algorithms.

\section{Charged Particle Tracking}

\begin{figure}[!ht]
\begin{center}
 \includegraphics[width=3.1in]{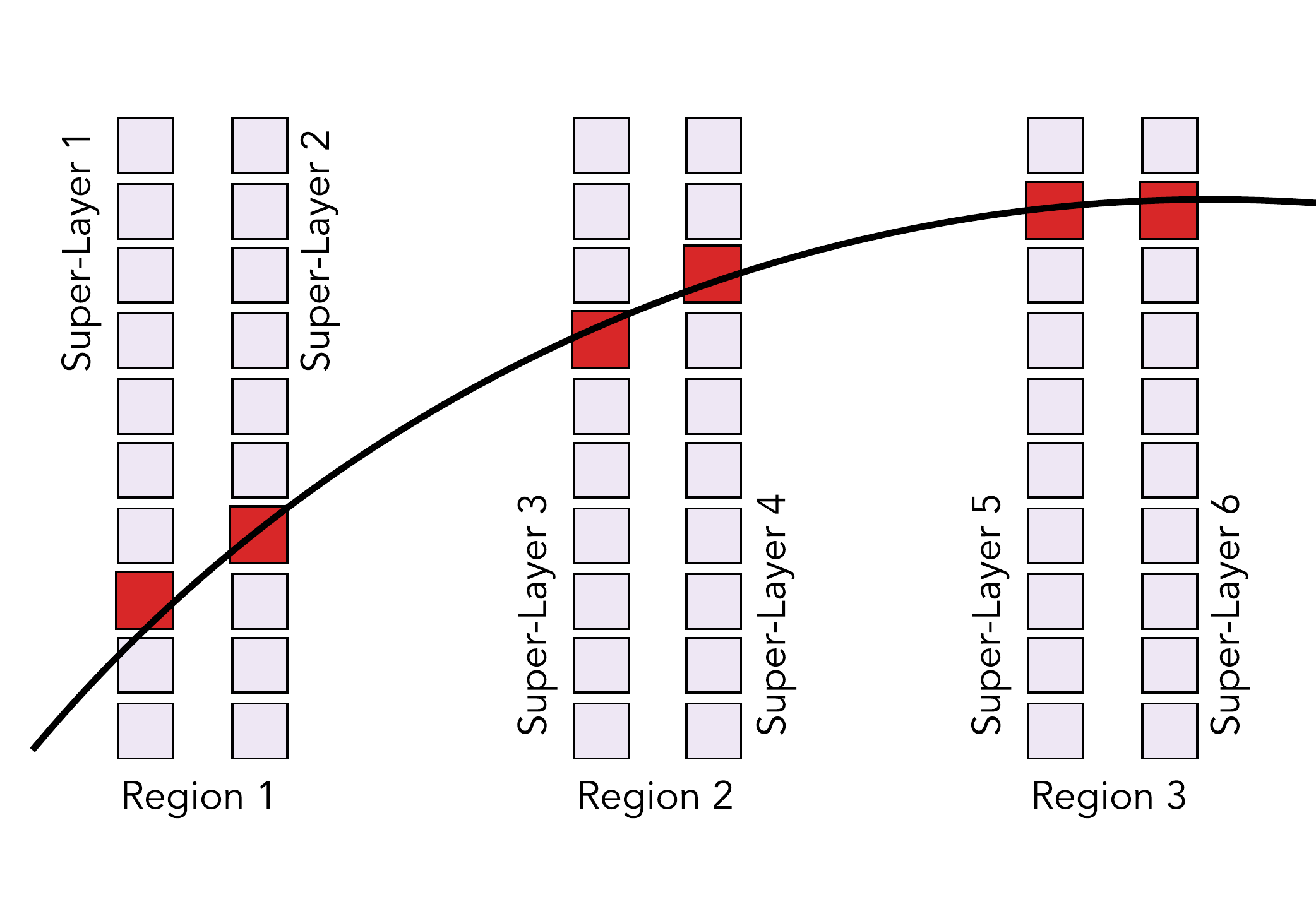}
 \includegraphics[width=3in]{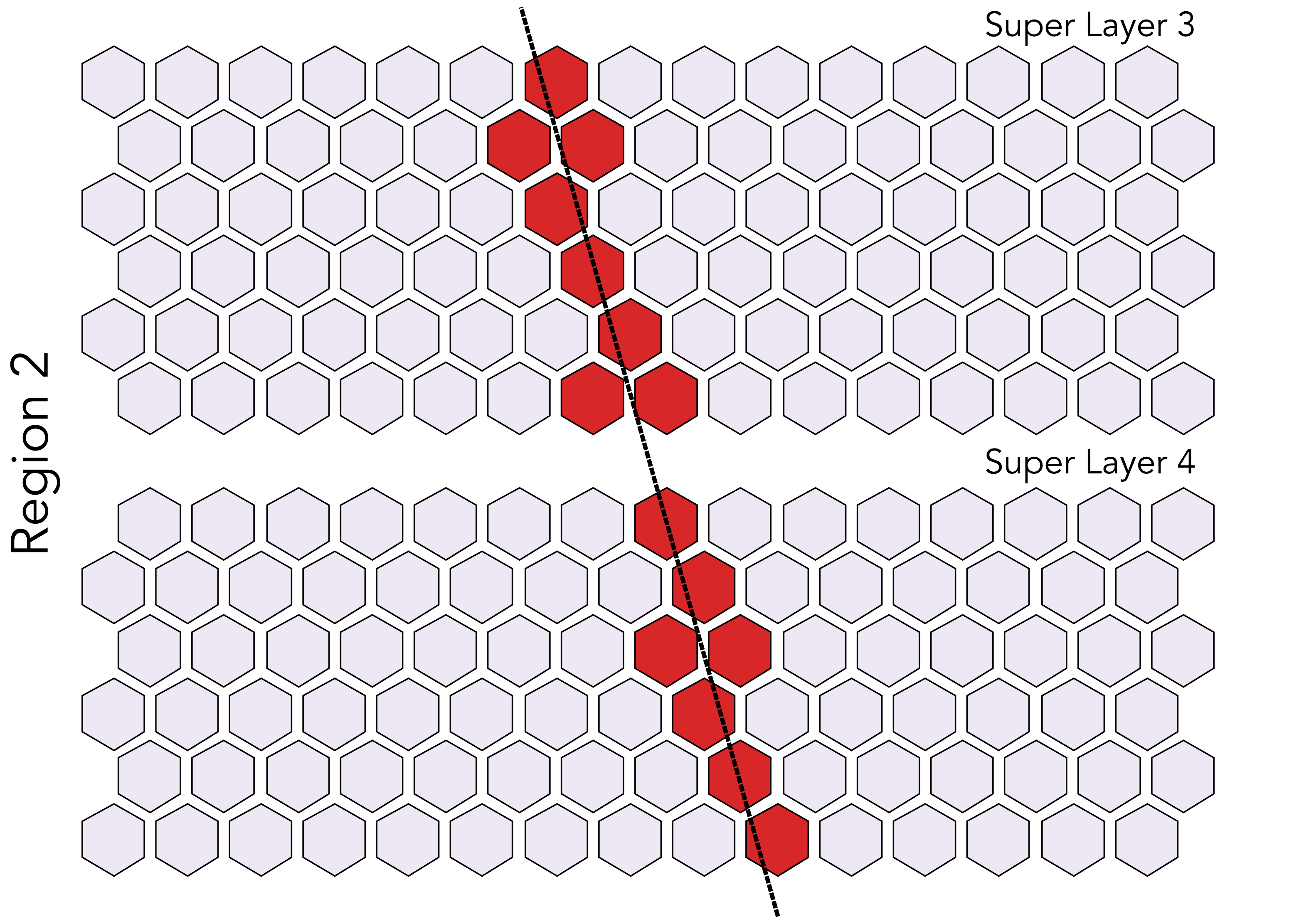}
\caption {Schematic view of signals generated in the drift chambers when a particle passes through. 
The segments in each super-layer are shown along the trajectory of the track (left panel), and view 
of the activated cells in the two super-layers of one region along the track trajectory (right panel).}
 \label{dc:side_view}
 \end{center}
\end{figure}

The CLAS12~\cite{Burkert:2020akg} forward detector is built around a six-coil toroidal magnet 
which divides the active detection area into six azimuthal regions, called ``sectors''. Each sector is 
equipped with three regions of drift chambers~\cite{Mestayer:2020saf} designed to detect charged 
particles produced by the interaction of an electron beam with a target. Each region consists of two 
chambers (called super-layers), each of them having 6 layers of wires. Each layer  in a super-layer 
contains 112 signal wires, making a super-layer a 6x112 cell matrix. The schematic view of one region 
is shown on Figure~\ref{dc:side_view} (right panel).

Particles that originate at the interaction vertex travel through the magnetic field and pass through all 
three regions of the drift chambers in a given sector are reconstructed by tracking algorithms. First, 
in each super-layer adjacent wires with a signal are grouped together into clusters (called segments), 
shown in Figure~\ref{dc:side_view}. The positions of these clusters (segments) in each super-layer 
are used to fit the track trajectory to derive initial parameters, such as momentum and direction. 
After the initial selection, good track candidates are passed through Kalman filter~\cite{Kalman1960} 
 to further refine measured parameters.

\begin{figure}[!ht]
\begin{center}
 \includegraphics[width=6.2in]{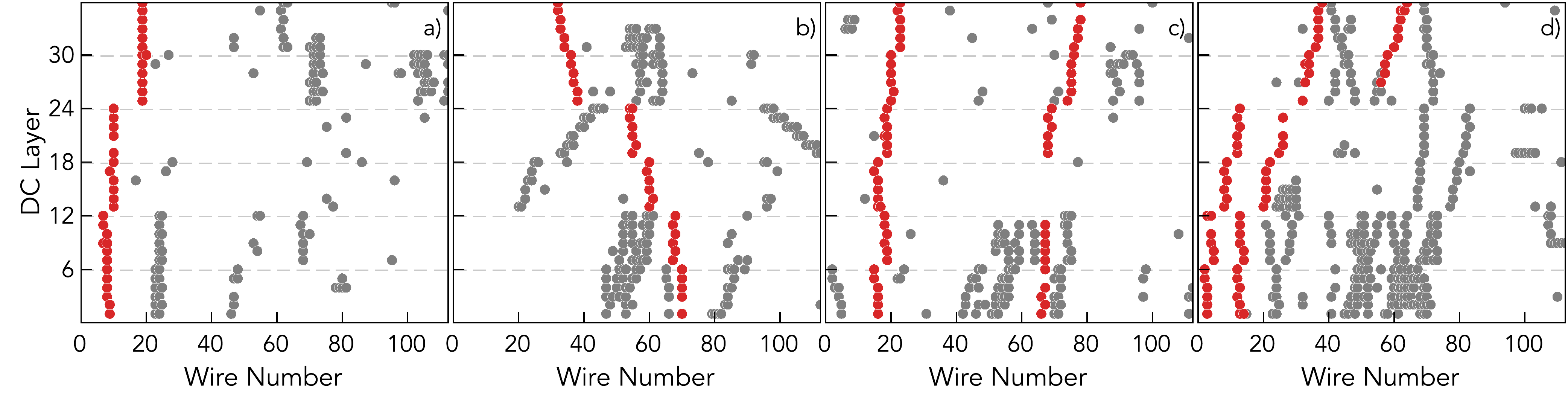}
\caption {Example of signals in drift chambers for four events. Each plot represents one sector with a 
36x112 matrix of wires. Background hits (in gray) are shown along with the hits of reconstructed 
identified tracks (in red). Dashed lines represent boundaries between super-layers.}
 \label{dc:events_sector}
 \end{center}
\end{figure}

For each beam-target interaction or ``event'', drift chambers produce many segments, some belonging 
to a track and some are background, or partial trajectories of low momentum tracks. In Figure~\ref{dc:events_sector} 
drift chamber signals in one sector are shown for four different  events, in each sector data are hits 
represented as a 36x112 matrix (36 layers and 112 wires per layer), showing all hits including those that 
were determined to be part of a track. 

Due to inefficiencies in the drift chambers, it is possible to have one missing segment along the trajectory 
of the particle, and the track has to be reconstructed using only 5 segments. An example of a 5-segment 
track is shown in Figure~\ref{dc:events_sector} c), where super-layer 3 does not have any segment detected. 
For these types of tracks, candidates have to be identified from a large number of combinatorics consisting 
of all combinations of clusters that form 5-segment candidates. 

Tracking is computationally intensive and makes up $80\%-90\%$ of the total CLAS12 event processing 
time (depending on background conditions and track multiplicity). The procedure of finding tracks from a 
list of track candidates is where we found AI can provide real benefits. Such benefits include improved 
accuracy in identifying good tracks and improved data processing speed by significantly reducing the 
number of candidates that have to go through initial fitting and then through the Kalman filter. AI can also 
help in identifying 5 super-layer track candidates.

\section{Machine Learning}

Machine Learning is used to train a neural network to recognize good tracks
 from a large list of candidates, constructed using one cluster from each super-layer.
 Two neural networks were developed: a classifier that can identify a good track from  
 6-segment track candidates and an auto-encoder that can take a list of 5-segment tracks 
 and turn them into 6-segment track candidates by adding a pseudo-segment. The composed 
 6 super-layer track candidates can be processed with the classifier network to identify and 
 isolate``good'' track candidates.

 \subsection{Track Classifier}
 
\begin{figure}[!ht]
\begin{center}
  \includegraphics[width=4.5in]{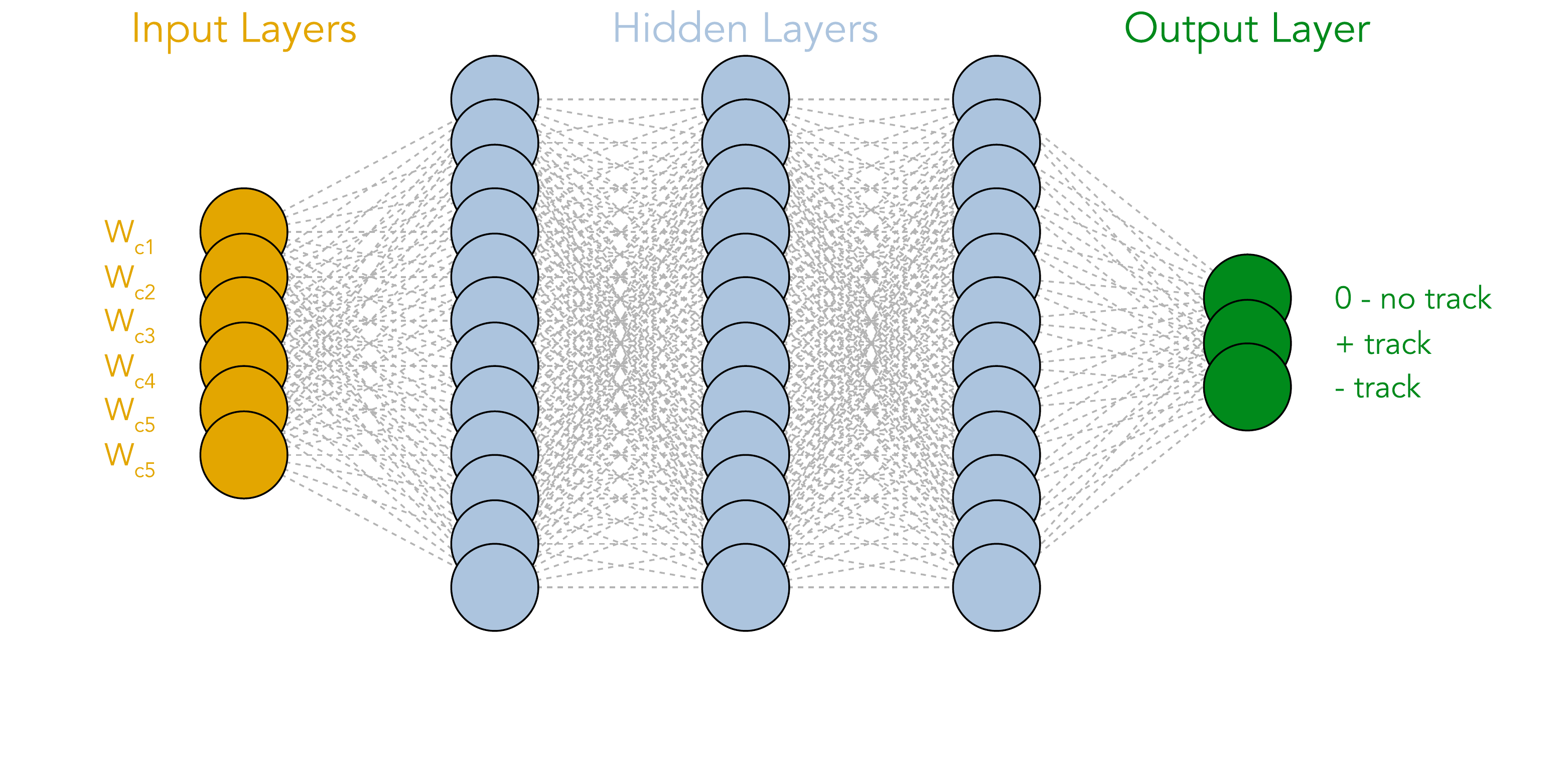}
\caption {Architecture of Multi-Layer Perceptron used for track classification. Network has 6 input nodes,
corresponding to average wire positions of the segment in each super-layer, three hidden layers with 12 nodes 
in each and 3 output nodes.}
 \label{mlp:architecture}
 \end{center}
\end{figure}

 To determine what type of machine learning model works best with the CLAS12 drift chamber data, 
 we investigated different types of models  \cite{Gavalian:2020oxg}, including Convolutional Neural 
 Network (CNN) , Extremely Randomized Trees (ERT) \cite{scikitlearn-extratreesclassifier}, and 
 Multi-Layer Perceptron (MLP) \cite{scikitlearn-mlpclassifier}. The study showed that Multi-Layer 
 Perceptron was best suited for the CLAS12 reconstruction needs (based on inference speed and accuracy). 
 The implemented architecture is shown in Figure~\ref{mlp:architecture}, where an input layer with 6 
 nodes is used (each node representing the average wire position of the segment in super-layer) and 3 
 output nodes for the classes ``positive track'', ``negative track'' and ``false track''.
 
The network is trained using a sample of tracks reconstructed by the conventional algorithm, selected with 
$\chi^2$ cuts to retain the highest quality ones, which are fed to the network with their respective labels 
(i.e. positive or negative tracks). For false tracks, a combination of segments (6-segments forming a track 
candidate) that was not identified as a track by the conventional algorithm is chosen.
 
 \subsection{Corruption Auto-Encoder}
 
\begin{figure}[!ht]
\begin{center}

\includegraphics[width=4.5in]{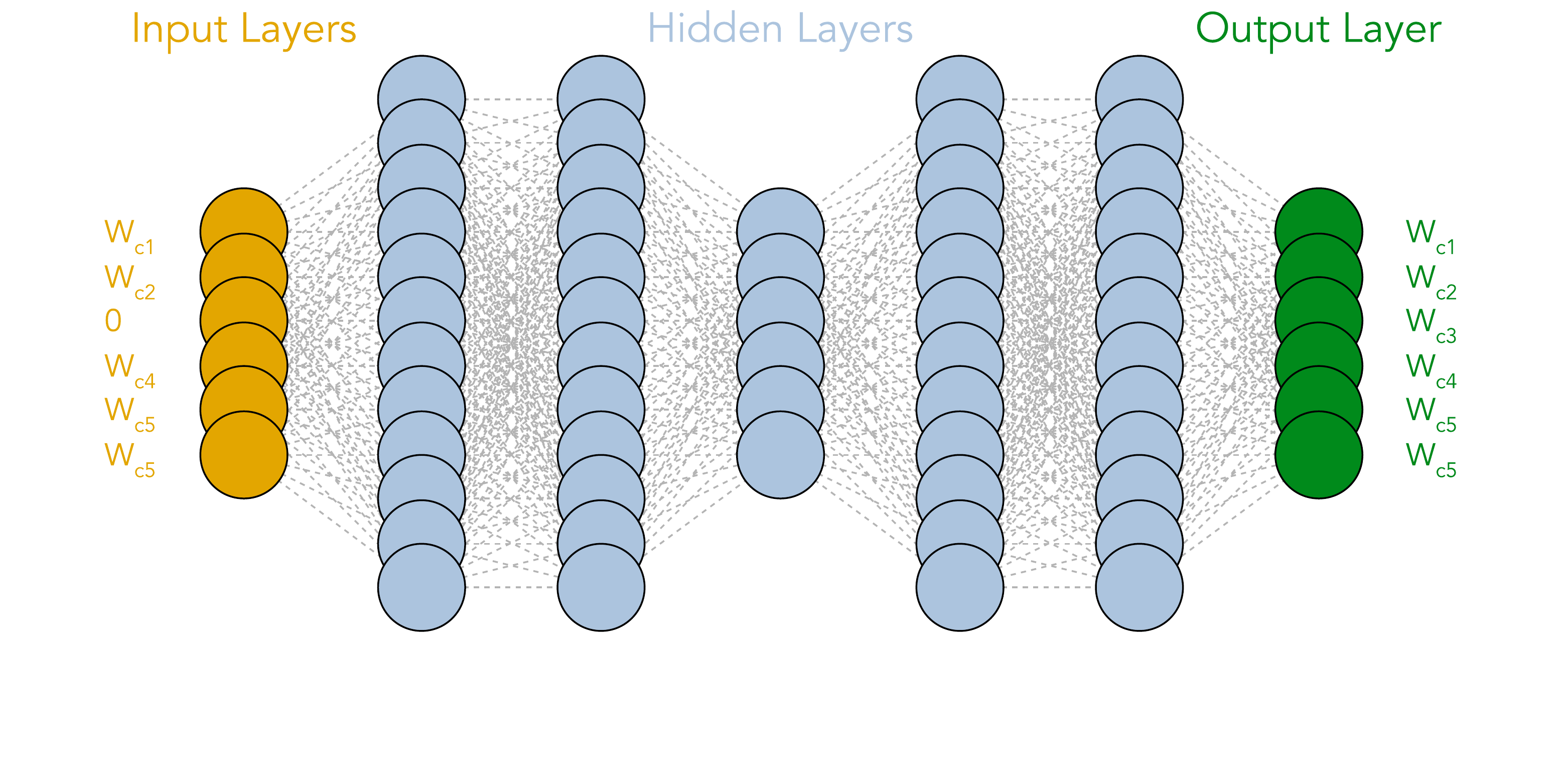}
\caption {Corruption-recovery auto-encoder architecture with 6 input nodes representing track segments 
mean wire values with one of the values is set to 0, and 6 output nodes with the correct value for the node 
that has 0 in the input. }
 \label{autoencoder:architecture}
 \end{center}
\end{figure}

A second neural network was developed to fix the corruption in possible track candidates due to 
inefficiencies of drift chambers. This network was used to identify track candidates which have one of 
the segments missing. We used an auto-encoder architecture to implement the corruption-recovery 
neural network \cite{Gavalian:2020xmc}. The structure of the network can be seen in 
Figure~\ref{autoencoder:architecture}, with 6 input nodes and 6 output nodes.

To train the corruption auto-encoder network the same sample used for the classifier training was used.
The output for the network was set to the good track parameters (where all 6 segments have non-zero values) 
and the input was modified by setting one of the nodes (randomly) to zero. The network learns to fix the node 
containing zero, by assigning it a value based on the other 5 segment values. 

 \begin{figure}[!ht]
\begin{center}
\includegraphics[width=6.0in]{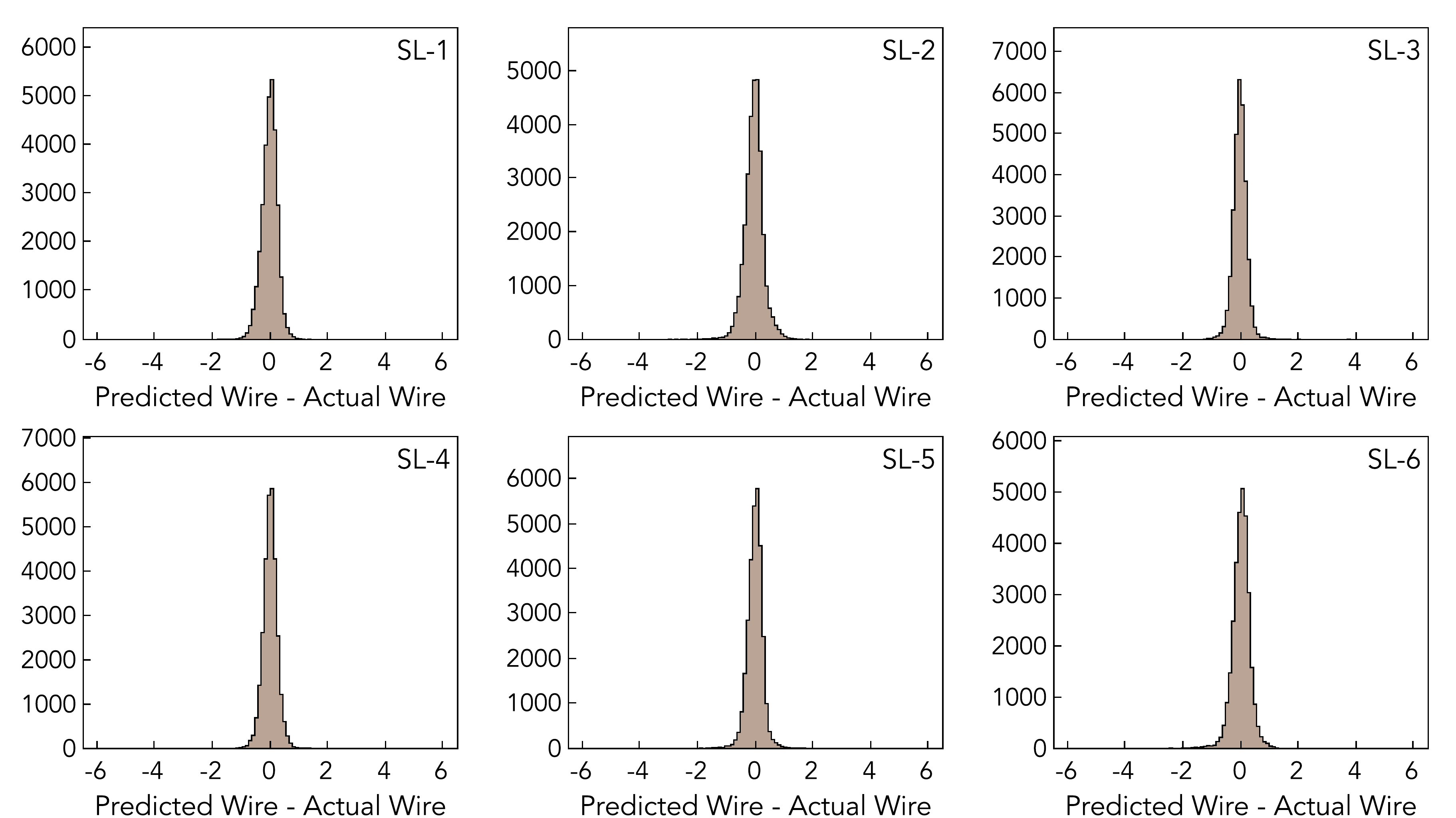}
\caption {Performance of corruption recovery auto-encoder foe each of six super-layers. The difference between predicted and actual 
wire position is shown for all super-layers. Average accuracy of wire position prediction is  $0.36$ wire.}
 \label{autoencoder:performance}
 \end{center}
\end{figure}

The test results of the trained network are shown in Figure~\ref{autoencoder:performance}, where the 
difference between the true value of the segment position and the one reconstructed by the network is 
plotted, showing a reconstruction accuracy of $0.36$ wires. In Figure~\ref{autoencoder:architecture} 
this difference is shown for each super-layer that was corrupted in the input. As can be seen, 
the performance of the network is uniform across all super-layers.

\section{Implementation of the Neural Network in CLAS12 software}

The models described in the previous sections were implemented in the CLAS12 tracking software. 

\subsection{Track Identification Workflow}

 Track identification consists of two phases, programmed to be done in two passes. In the first pass 
 over the data, signals from each sector of drift chambers are analyzed to create a track candidate list, 
 each consisting of 6 segments. The resulting track candidates are evaluated by the classifier neural 
 network and are assigned a probability of being either a positive or negative track. The list of track 
 candidates is sorted by probability and passed to another algorithm that is responsible for removing 
 tracks that have a lower probability of being a "good" track and have clusters that are shared with a 
 higher-probability candidate. 

In this procedure, the algorithm iterates over-track candidates sorted according to the probability of 
being a good track. Iteration starts at position number 2 and runs to the end of the list. Candidates that 
share a cluster with candidates at position number 1 are removed from the list. Track candidate at position 
1 is moved from the track candidate list into the identified track list. This procedure is repeated until 
there are no track candidates left in the candidate list.

 \begin{figure}[!h]
\begin{center}
 \includegraphics[angle=90,width=1.1in]{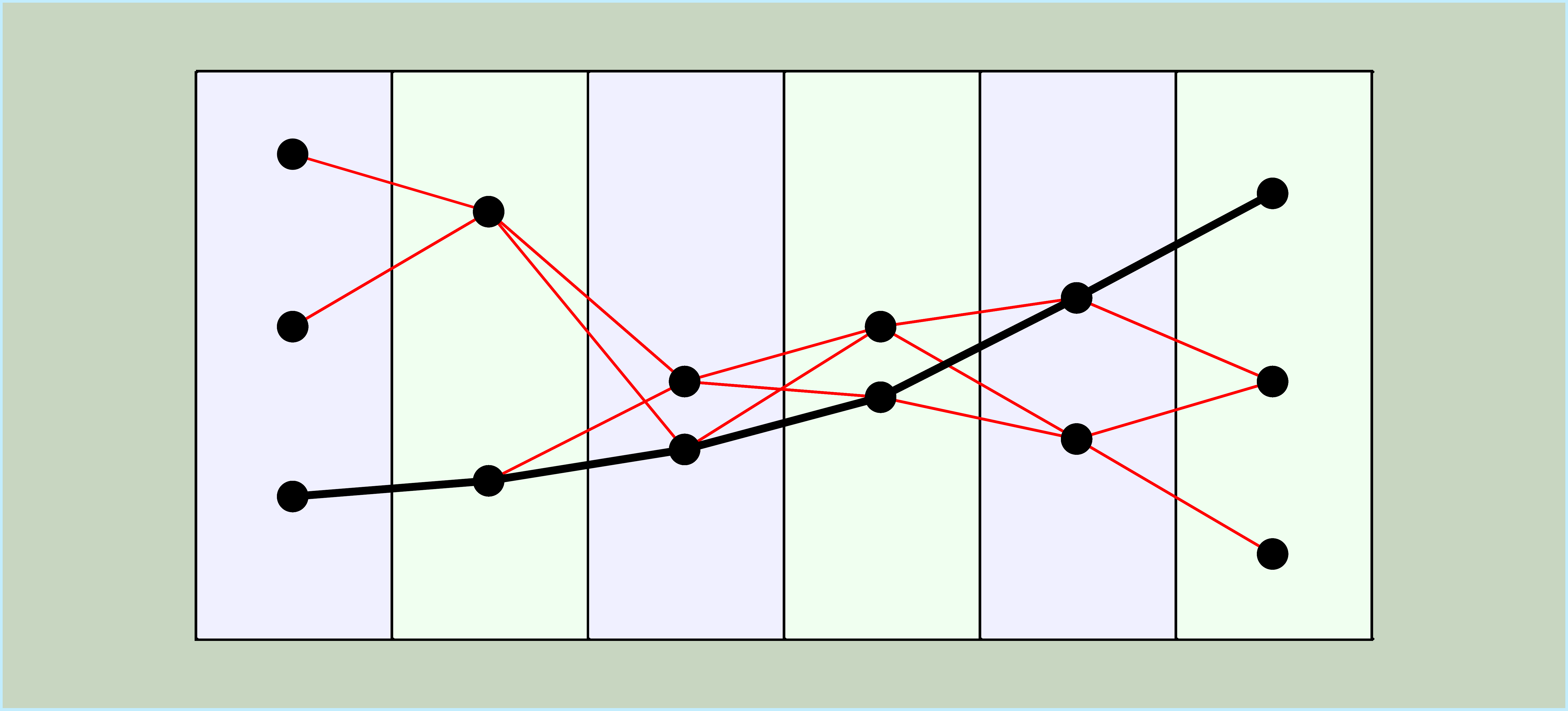}
  \includegraphics[angle=90,width=1.1in]{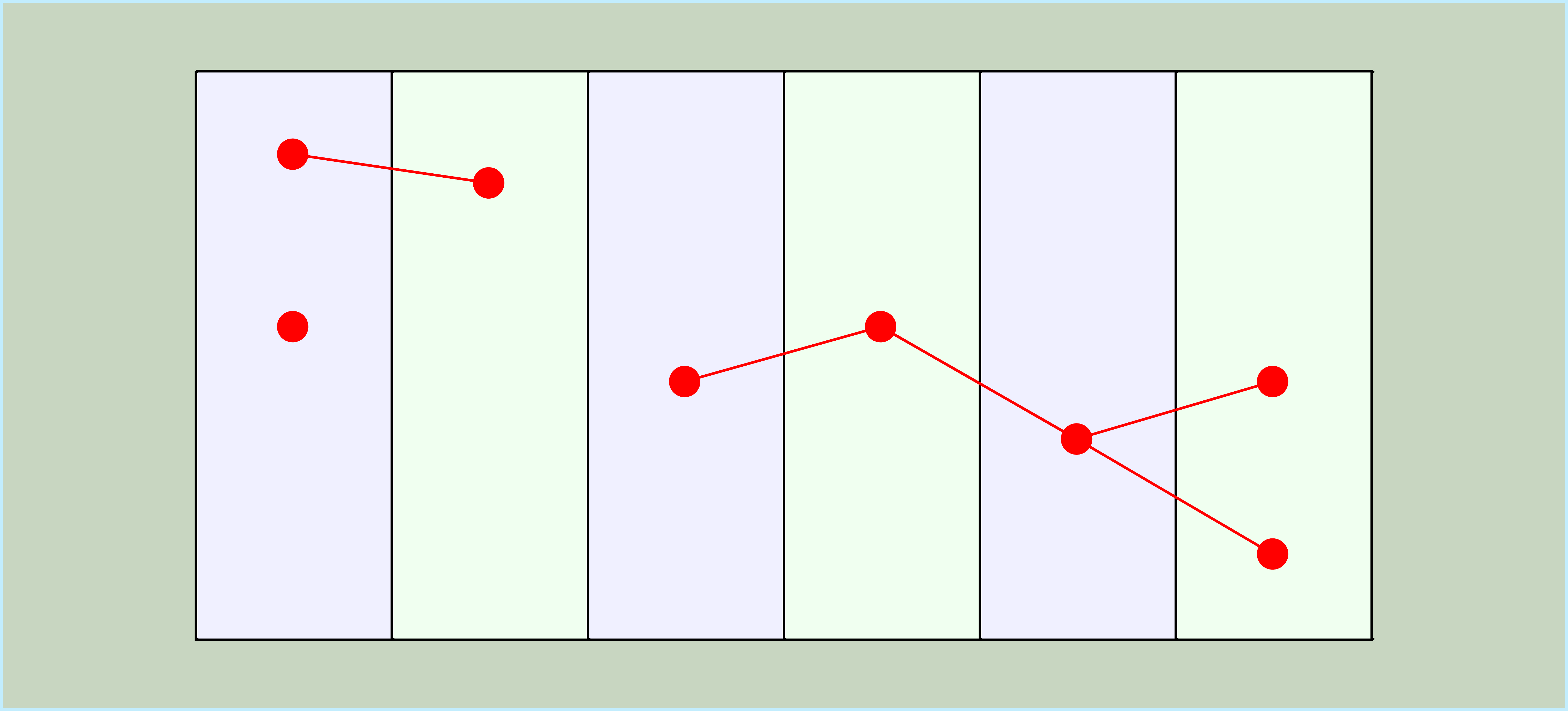}
    \includegraphics[angle=90,width=1.1in]{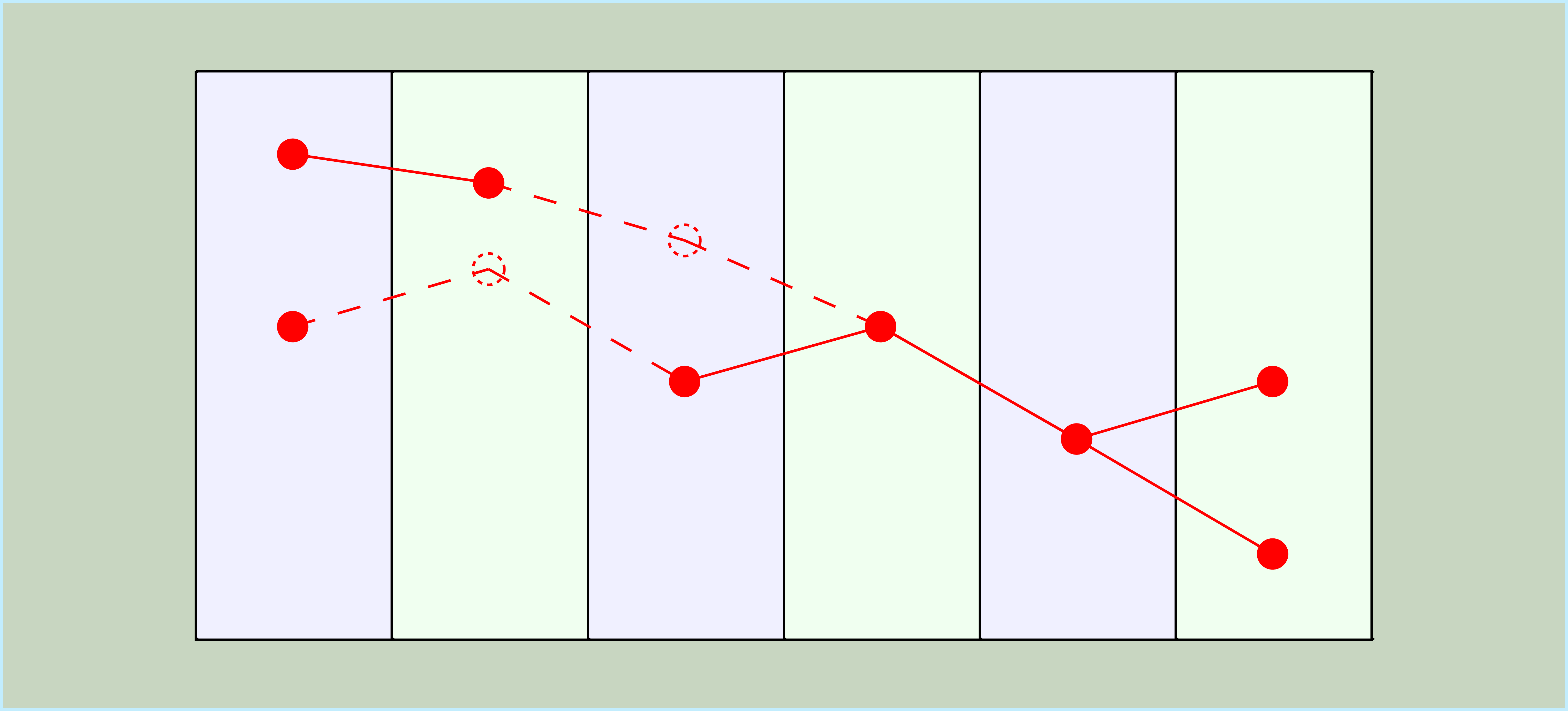}
      \includegraphics[angle=90,width=1.1in]{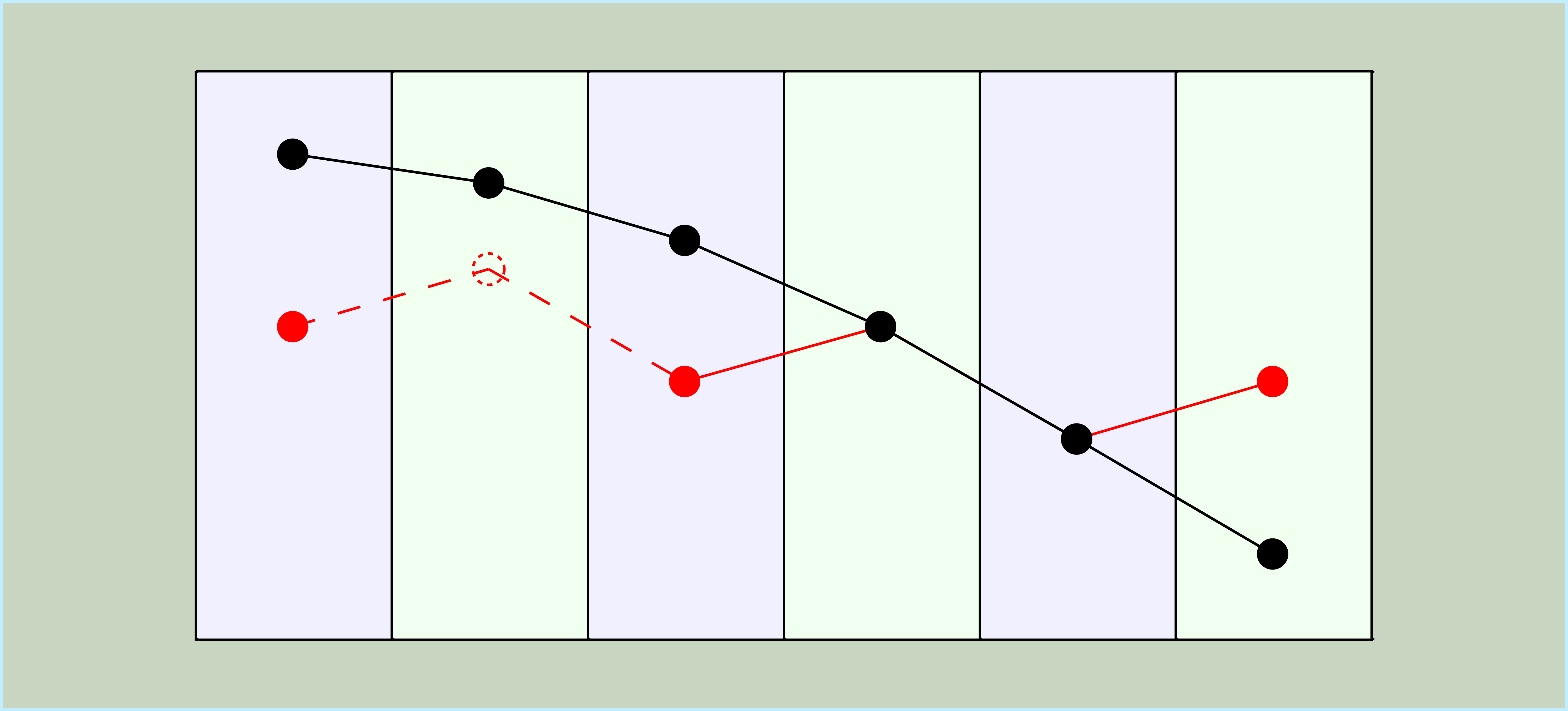}
            \includegraphics[angle=90,width=1.1in]{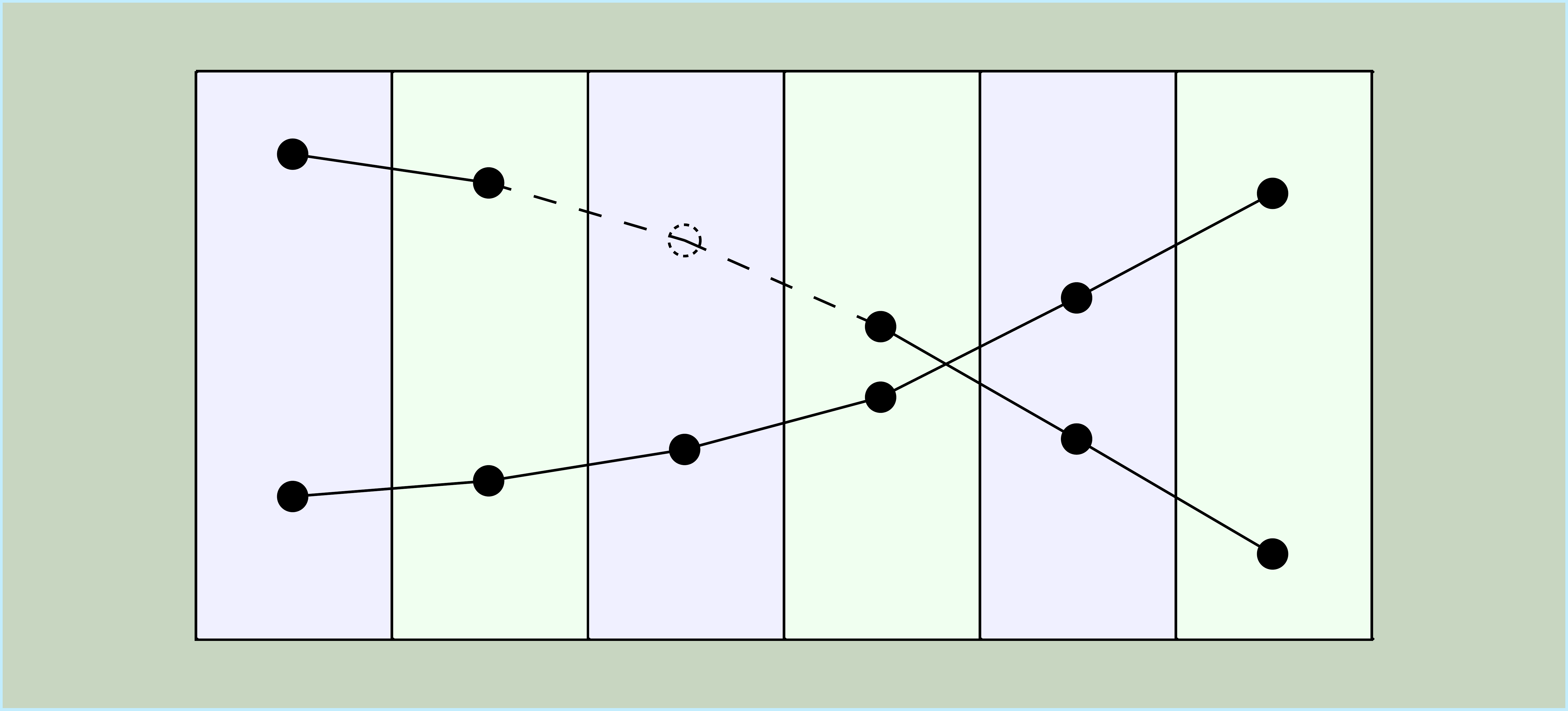}
            
\caption {Stages of Neural Network track identification procedure. 1) identifying 6 super-layer tracks. 2) 
removing all hits belonging to an identified track and constructing 5 super-layer track candidates. 
3) generating pseudo-clusters for 5 super-layer track candidates using corruption fixing auto-encoder. 
4) identify good track candidates from the list of 6 super-layer (one of the super-layers is a pseudo-cluster) 
track candidates. 5) isolate both identified (6 super-layer and 5 super-layer) tracks  for further fitting with 
Kalman-Filter.}
 \label{network:procedure}
 \end{center}
\end{figure}

The second stage of track identification starts by constructing a list of track candidates with combinations 
of 5 clusters out of 6 from all existing clusters (one per super-layer). The candidates that share a cluster with 
tracks identified at the first stage of classification are removed from the list. For each track candidate with a 
missing cluster in one of the super-layers, a pseudo-cluster is generated using the Corruption Auto-Encoder 
Network and the missing super-layer cluster are assigned the inferred value, hence turning all track candidates 
to 6 cluster track candidates. The cured (or fixed) track candidate list is finally passed to the track classifier 
module described above, which evaluates the list isolating candidates with the highest probability of being a 
good track. 

\subsection{Implementation in reconstruction software}

The CLAS12 reconstruction software framework is a Service Oriented Architecture platform implemented in Java (CLARA~\cite{Gyurjyan:2011zz}).
The reconstruction software consists of several microservices, each responsible for processing data from one
detector \cite{Ziegler:2020gsr}. The reconstruction procedure for some of the detector components can also 
be broken down into smaller logical microservices to add some flexibility in changing the implementation of 
the small parts and provide alternative reconstruction procedures for some of the components. Reconstruction 
of tracks in drift chambers is a complex task and consists of several parts.

The first stage of the process, called clustering service, is to isolate clusters from the hits in drift chambers. 
Once the clusters are isolated, track candidates are formed from all combinations of 6 segments. Once 6 
segment tracks have been identified, the remaining segments are then used to form 5 segment combinations. 
These two steps are known as track-finding or seeding. Both 6 and 5 segment candidates are fitted defining 
the hit positions from the wire coordinates (hit-based tracking), resulting in the first list of reconstructed tracks.

In a second stage of the tracking code (time-based tracking), the tracks identified at the previous stage are 
refitted with a Kalman Filter algorithm, which uses drift time information to refine the hit positions. This 
produces the final track list.

\begin{figure}[!ht]
\begin{center}
 \includegraphics[width=6.0in]{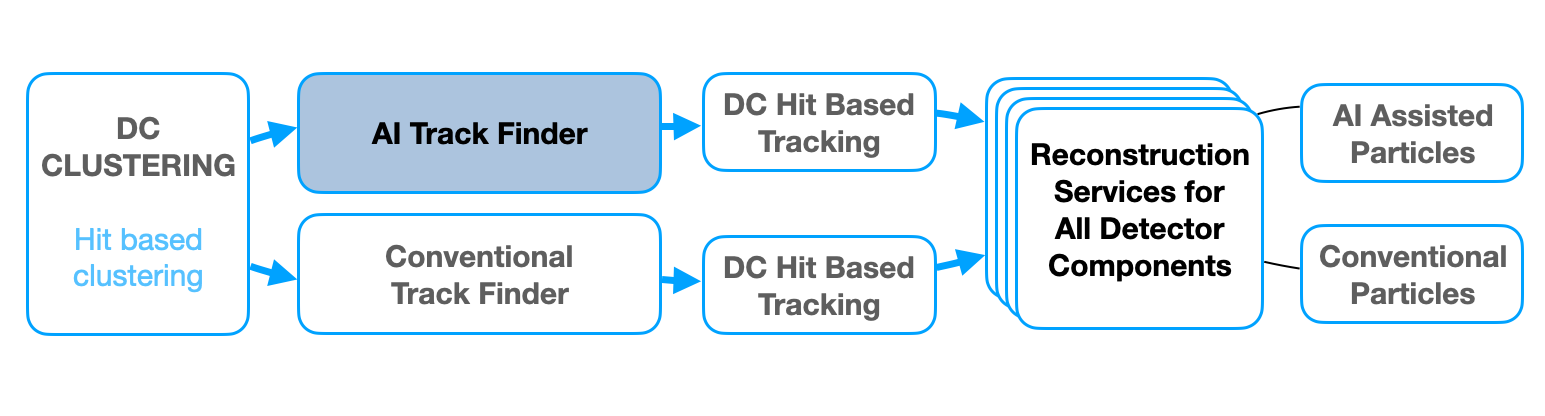}
\caption {Diagram of the tracking workflow with Artificial Intelligence included. The Workflow is split into 
two parallel branches, one where track-finding is done by the conventional algorithm, and one where only 
AI-isolated tracks are fed to the following stages.}
 \label{recon:diagram}
 \end{center}
\end{figure}

In order to implement our neural network into the reconstruction workflow, we designed two parallel 
branches in the reconstruction code where we run two algorithms to identify good tracks from the 
candidate lists, one based on the conventional algorithm and one based on the neural network. The two 
algorithms store their track suggestions in separate data structures and pass them to the next stage where 
track parameters are reconstructed by a conventional tracking algorithm, first by hit-based fitting and then 
using the Kalman filter. This approach lets us have two parallel outputs from the tracking code that enables 
a detailed comparison of the performance of each method.

\subsection{Software Packages}

As mentioned above, CLAS12 reconstruction is implemented in Java. The final implementation of the  
track classifier was therefore in Java for easy integration into the reconstruction workflow. The initial tests 
and prototyping were done using the Keras/TensorFlow~\cite{keras-website} (python) package. The final 
implementation was based on the DeepNetts \cite{Sevarac.Z} community edition library (in native Java) 
used for both the track candidate classifier and corruption-recovery auto-encoder. DeepNetts is a lightweight
 library with minimal dependencies, which makes it ideal for providing portable code that can be used on a variety 
 of platforms without the need of installing a large number of platform-dependent packages (like in the case of
  keras/tensorflow). The inference procedure was implemented using Efficient Java Matrix Library (EJML)~\cite{ejml:2021}, 
  which is optimized for speed and is thread-safe, matching the requirements of the CLAS12 reconstruction software.

The analysis and data visualization for this article was done using GROOT data visualization package~\cite{groot-github} 
developed for CLAS12 software infrastructure (in Java) and is included in the Java data analysis library for high energy 
physics Jas4pp~\cite{Chekanov:2020bja}.

\section{Analysis of Track Reconstruction with AI}

After implementing the track identification service in the CLAS12 
reconstruction software, the outputs from the conventional tracking 
algorithm and AI-assisted tracking algorithm were analyzed event by 
event to assess the improvements in tracking. 
 
 \subsection{Particle Reconstruction efficiency}
 
 The Neural Network for track classification was trained on experimental 
 data after it was processed with conventional tracking. Tracks that had ``good'' 
 fit quality and were tracked back to the target location were used as training 
 samples for both the MLP classifier and Auto-Encoder corruption-recovery network. 
 Performance was assessed on data recorded at CLAS12 production settings, 
 specifically with 45 nA electron beam impinging on a 5 cm-long hydrogen target 
 for an instantaneous luminosity close to $0.6\times10^{35} cm^{-2} s^{-1}$.

 \begin{figure}[!ht]
\begin{center}
  \includegraphics[width=6.5in]{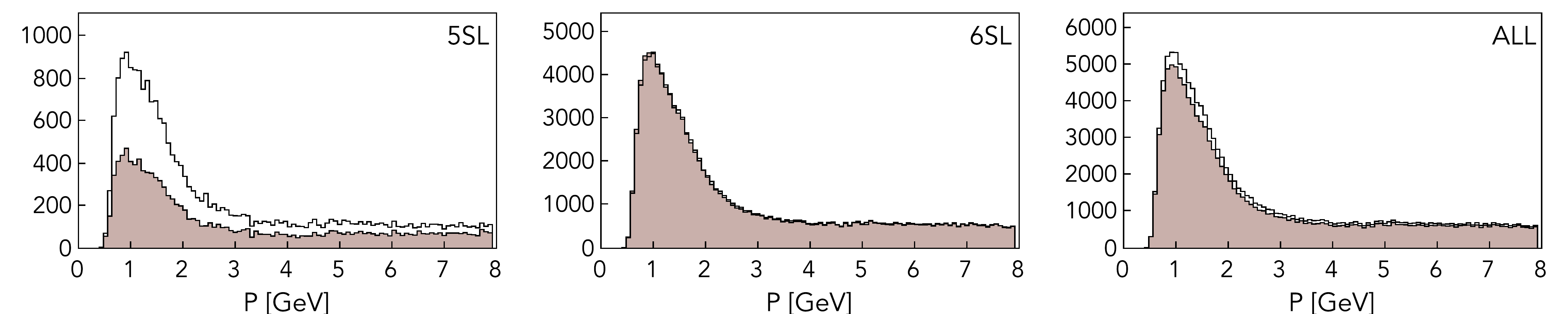}
  \includegraphics[width=6.5in]{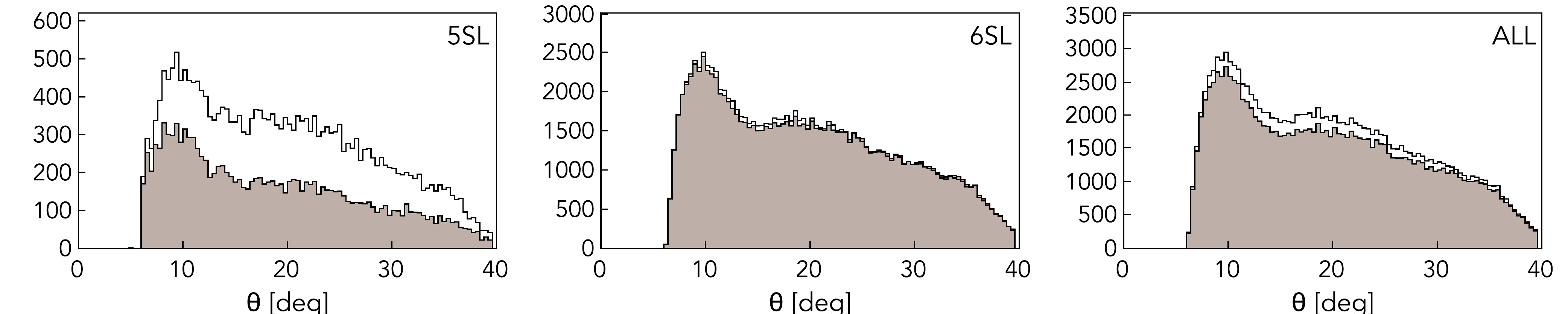}
    \includegraphics[width=6.5in]{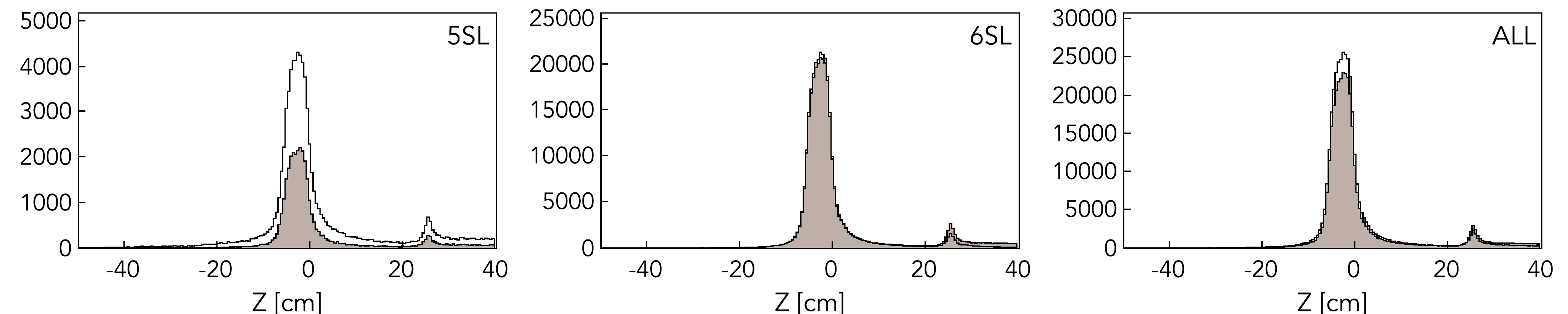}
\caption { Comparison of number of tracks reconstructed with the conventional 
algorithm (filled histograms) vs AI-assisted tracking code (open black outline 
histograms) as a function of momentum, polar angle and particle interaction vertex. 
The comparison is shown for 5 super-layer, 6 super-layer tracks (left two columns), 
and the total number (right column).}
 \label{track:efficiency}
 \end{center}
\end{figure}

The results are shown in Figure~\ref{track:efficiency}, where the dependence of the number 
of ``good'' reconstructed negatively charged tracks is shown as a function of particle 
momentum (top row), polar angle in laboratory frame (middle row), and interaction vertex 
(bottom row). The reconstructed distributions from conventional tracking are plotted with 
filled histograms and the tracks reconstructed using assistance from AI are plotted with 
solid lines. As can be seen from the figure, there is a large gain in the number of reconstructed 
tracks with the 5 super-layer configurations compared to the full 6 super-layer tracks. Typically 
for nominal 45 nA experimental data increase in track efficiency averages about $3\%-6\%$ 
for 6 super-layer tracks and to $70-120\%$ for 5 super-layer tracks. In conventional data 
reconstruction, tracks that are identified with 5 super-layers usually comprise about $10\%$ 
of all reconstructed tracks, and a significant increase in identification of such tracks leads to 
an overall tracking efficiency increase of $10\%-15\%$. 
 
 \begin{table}[!h]
 \begin{center}
 \begin{tabular}{|l|c|c|c|c|}
 \hline
 Track Configuration & Conventional & AI Assisted & Gain & Relative \\
 \hline
 \hline
 6 Super-Layer & 242,145 & 256,175 & 14030 & 1.0579 \\
 5 Super-Layer & 24,155 & 52,839 & 28684 & 2.1875 \\
 All & 267,339 & 309,058 & 51719 & 1.1561 \\
 \hline
 \end{tabular}
 \end{center}
 \caption{Summary of reconstructed tracks and gain with assistance from Artificial Intelligence algorithm.}
 \label{tbl:summary}
 \end{table}
 
The comparison of 5 and 6-segment track statistics and their relative gain is 
summarized in Table~\ref{tbl:summary}. As can be seen from the table, the 
gain in 6-segment tracks is about $5.7\%$ but with a significant gain in 5 super-layer tracks 
the overall gain in reconstructed tracks elevates to $>15\%$. These results are intuitive 
since the number of track candidates is composed of 5 super-layers, assuming a uniform 
number of clusters in the six super-layers is significantly higher than 6 super-layer track 
candidates, and in our tests AI performs better in choosing the right combination with 
increasing combinatorics.
 
\subsection{Luminosity Dependence}

As discussed in the previous section, AI is more efficient than the conventional track-finding 
algorithm in identifying good candidates from a large pool. One would expect that, if the number 
of combinations decreases, the efficiency of the conventional track selection should approach the 
efficiency of AI-assisted track identification. Similarly, when the number of combinations 
increases the gain of AI over the conventional algorithm should increase. Based on this we 
expect AI to perform better in higher luminosity, and higher background settings. To evaluate the 
AI-assisted tracking efficiency dependence on the background we analyzed several different runs 
that were taken in different conditions (i.e. beam current) ranging from $5~nA$ to $70~nA$. 
To determine the tracking efficiency we first calculated the number of electrons ($N_e$) detected 
in the analyzed data sample (typically one run, or $2~hours$ of data taking) and then the number of 
positive and negative hadrons that were detected with the electron inclusively ($N_{h^+e}$ and 
$N_{h^-e}$ respectively).

Then the efficiency for the data set was calculated as:

\begin{equation}
L_t^+ = \frac{N_{h^+e}}{N_e} , L_t^- = \frac{N_{h^-e}}{N_e} 
\end{equation}

where $L_t^+$ is the efficiency of positively and $L_t^-$ is the efficiency of negatively 
charged particles respectively. In order to estimate the charged-particle reconstruction efficiency 
as a function of the beam current, the multiplicity, $L_t^{+/-}$, is fitted with a linear function:
\begin{equation}
L_t^{+/-} = a + b\times I 
\end{equation}

Here $a$ and $c$ are the fit parameters and $I$ is the beam current. Then it was assumed that the 
reconstruction efficiency, $E=1$ at $I=0$ nA:

\begin{equation}
E^{+/-} = 1 + c \times I 
\end{equation}

with $c=\frac{b}{a}$. The slope parameter $c$ represents the variation of the reconstruction 
inefficiency per $nA$~\cite{Stepanyan:2020bg}.
 
 \begin{figure}[!ht]
\begin{center}
 \includegraphics[width=3.0in]{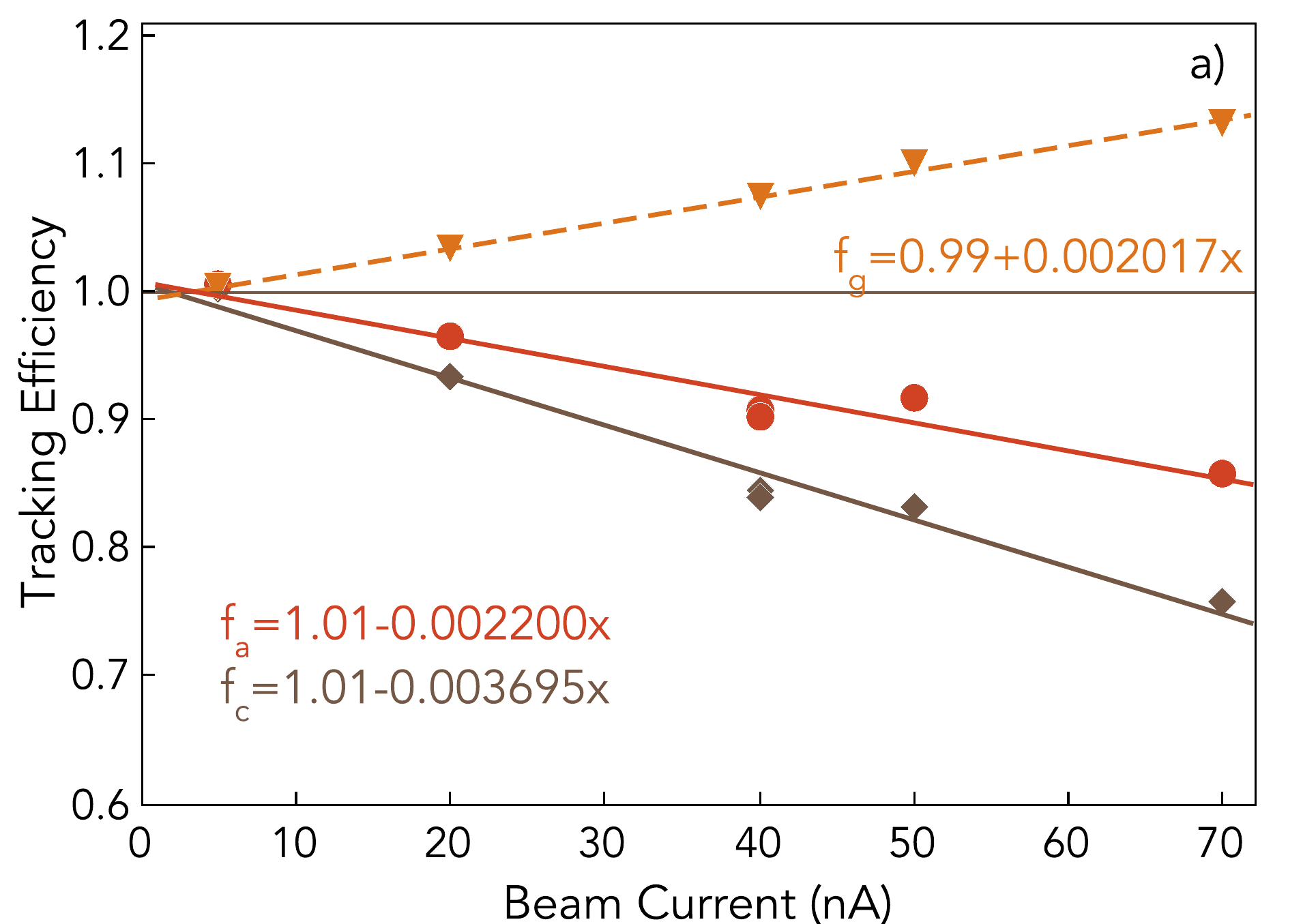}
 \includegraphics[width=3.0in]{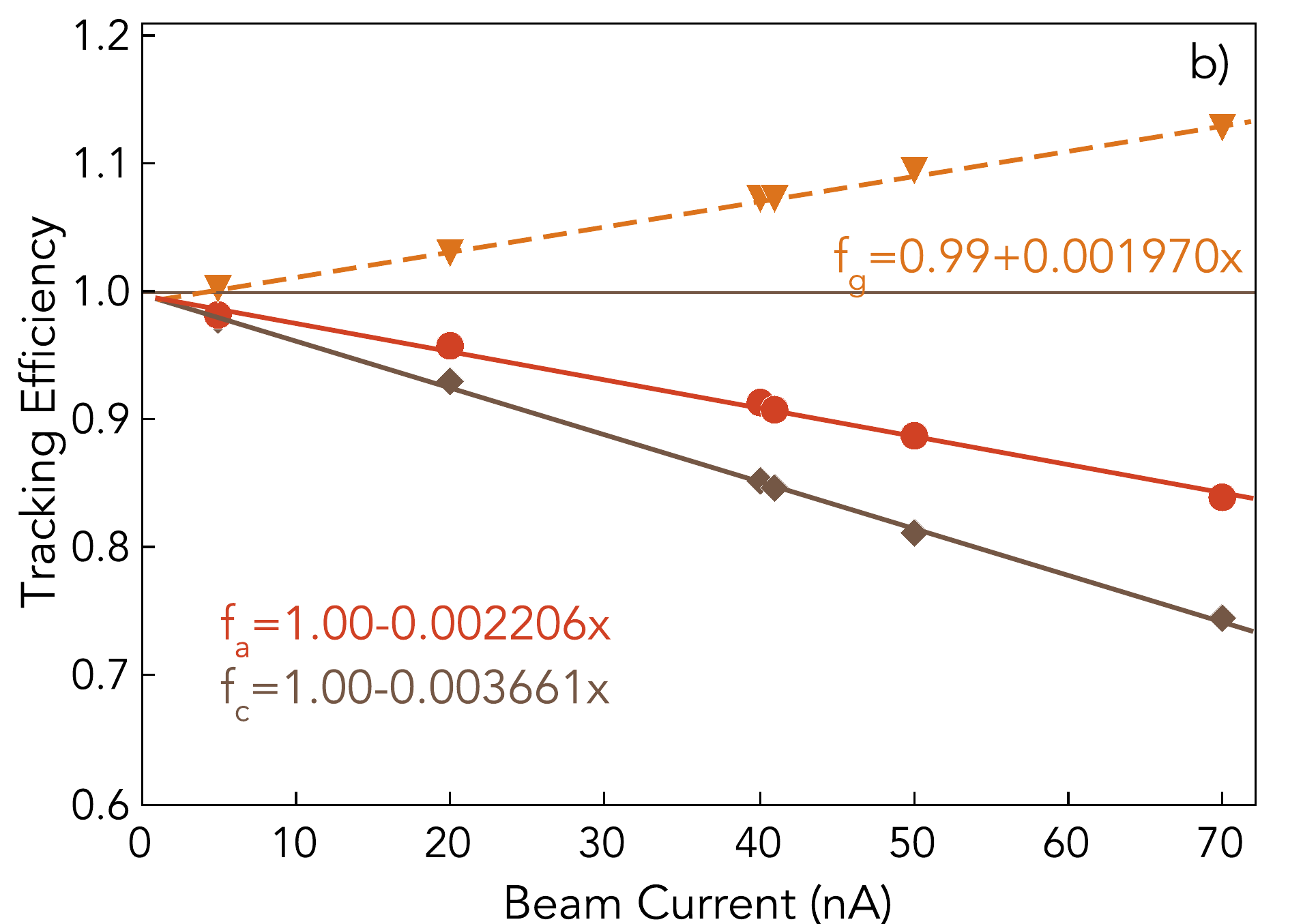}
\caption {Tracking reconstruction efficiency of the conventional algorithm (diamonds)
is compared to AI-assisted tracking efficiency (circles) for positively and negatively 
charged particles as a function of beam current (luminosity).   }
 \label{lumi:scan}
 \end{center}
\end{figure}

The comparison of tracking efficiency as a function of beam current (luminosity) can be 
seen in Figure~\ref{lumi:scan} where $E^{+/-}$ are shown for positively and negatively charged 
particles separately. AI-assisted tracking performs significantly better for any given luminosity 
(beam current) and the decrease of efficiency is much slower as a function of luminosity, $0.22\%$ 
per nA versus $0.40\%$ per nA for conventional tracking. This is expected and consistent with the 
assumption that with increased combinatorial background (increased number of track candidates 
to consider), AI performs better in choosing the best track candidate. We established that AI-assisted
tracking leads to more tracks reconstructed for any given beam current setting. The next thing to 
check is what is the impact of increased track reconstruction efficiency on physics analysis.

\subsection{Physics Impact}

To measure the implications of track reconstruction efficiency improvements on physics analysis, 
we considered two event topologies with two and three particles in the final state, respectively. 
The data for analysis were taken with $10.6~GeV$ electron beam incident on $5~cm$ liquid 
hydrogen target, with a beam current of $45~nA$. 
We selected events where an electron was detected in the forward detector, and then isolated 
events where there was an additional positively charged pion ($\pi^+$) along with an electron 
and no other charged particle. The second topology required two pions along with the electron, 
one positively charged and one negatively charged. The two chosen topologies are denoted by 
$H(e,e'\pi^+X)$ and $H(e,e'\pi^+\pi^-X)$. In both cases, there is a visible peak of a missing nucleon
in the missing mass distribution of the detected final state, which we can use to measure the impact 
of efficiency on physics outcome. 

 \begin{figure}[!ht]
\begin{center}
 \includegraphics[width=6.0in]{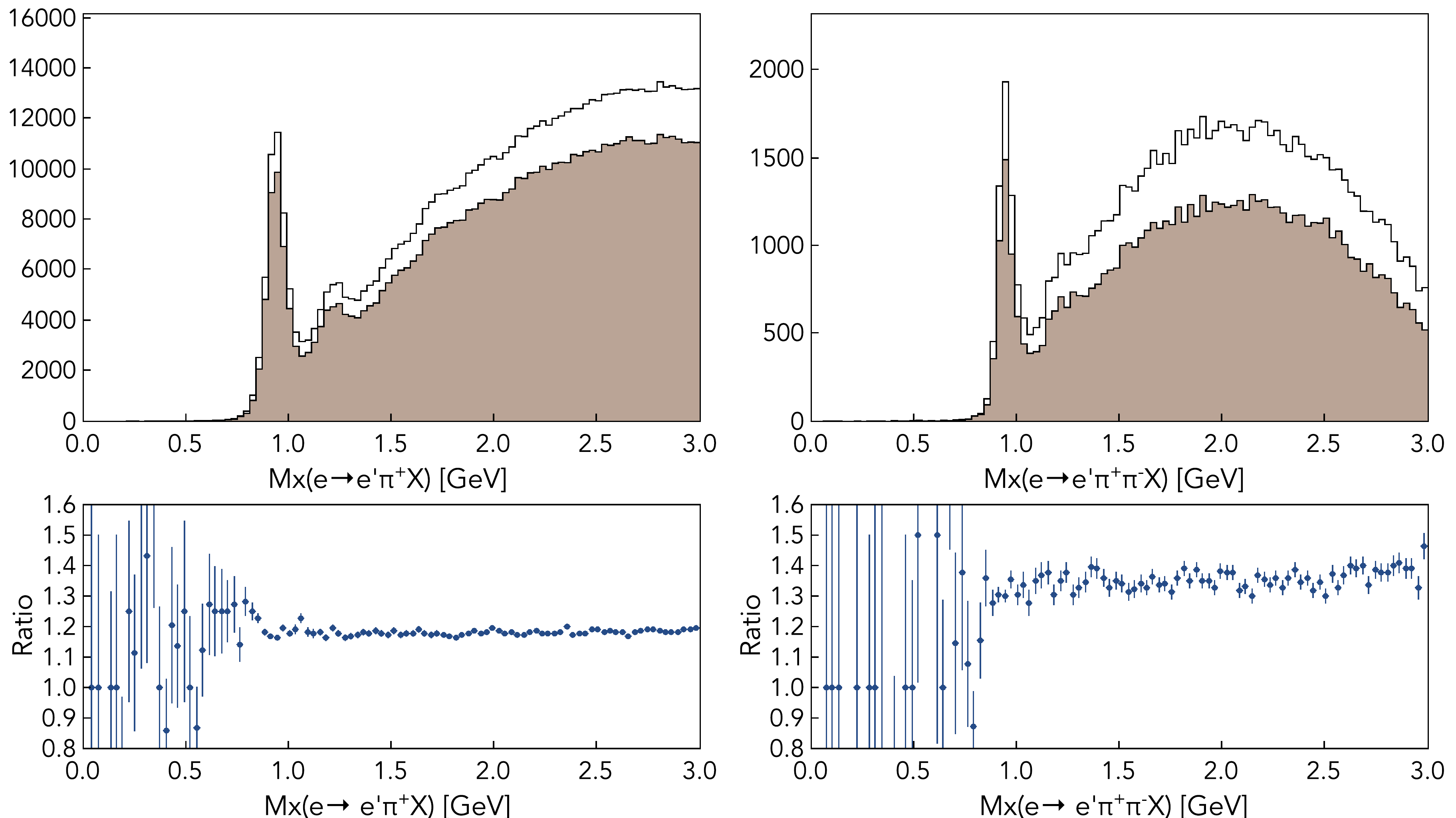}
\caption {Reconstructed missing mass distribution for $H(e,e'\pi^+X)$ and $H(e,e'\pi^+\pi^-X)$ 
reactions (top row) using the conventional track reconstruction algorithm (filled histogram) and  
AI-assisted track reconstruction (black line histogram). The ratios of the two histograms are shown 
on the bottom row. }
 \label{physics:outcome}
 \end{center}
\end{figure}

The distributions of missing mass for both final state topologies are shown in Figure~\ref{physics:outcome}, where the plots 
on the top row are missing mass of $H(e,e'\pi^-X)$ and $H(e,e'\pi^+\pi^-X)$, where the filled histogram is calculated from 
particles reconstructed by the conventional tracking algorithm, and the histogram with a black outline is the same distributions 
calculated from particles that were reconstructed using a suggestion from Artificial Intelligence. As can be seen from the figure, 
there is a significant increase in the number of events in the region of the nucleon peak for AI-assisted
tracking. The ratios of the two histograms (AI-assisted divided by conventional) can be seen on the bottom row of 
Figure~\ref{physics:outcome}. As can be seen from the figure the increase in statistics is uniform over the whole range of the 
missing mass indicating no systematic abnormalities for AI-assisted tracking. The ratio also indicates that there is an increase 
for the number of events of about $15\%$ for $H(e,e'\pi^+X)$ final state and $30-35\%$ for the $H(e,e'\pi^+\pi^-X)$
final state. Further studies show that improvements in statistics are larger for higher luminosity (higher incident beam current), 
which is consistent with our studies of increased efficiency of single-particle reconstruction.

\section{Summary}

In this paper we present results of the analysis of experimental data from the CLAS12 detector obtained with the assistance of Artificial Intelligence
to identify tracks from the hits in drift chambers. This work is based on two neural networks developed to classify track candidates from
given cluster combinations \cite{Gavalian:2020oxg} and to identify missing cluster positions in tracks that do not have complete 6 cluster configuration \cite{Gavalian:2020xmc}. After implementing these networks into the CLAS12 reconstruction workflow, AI was able to identify ``good'' track candidates 
and pass them to the tracking code to be analyzed in parallel to conventional algorithms that choose ``good'' track candidates iteratively considering all possible combinations. 
Our studies showed that AI-assisted tracking performs better than conventional track identification algorithm, and leads to track reconstruction efficiency increase of $10-12\%$ for beam current of 45 nA. The AI also performs better with increasing background (i.e. with increased incident beam current) and improves the efficiency loss from $0.37\%$ per nA to $0.22\%$ per nA.
This increased track reconstruction (identification) efficiency directly impacts the outcome of physics analysis, where it leads to an increase in statistics of 
$15\%-35\%$, depending on how many particles are detected in the final state. This has significant implications for the choice of experimental running conditions since with increased efficiency the required statistical significance can be reached in a shorter time by running at a higher beam current (luminosity). Already collected experimental data can be re-processed with the AI-assisted tracking
code which can increase the statistics available for physics analysis up to $35\%$. Therefore, both future experiments and already completed ones will benefit 
from this novel development.

Another important outcome of this development was a reduction in data processing times. Since track candidates were identified by AI, there were fewer marginal quality tracks picked to be analyzed and then later dropped due to non-convergence of Kalman filter, leading to tracking code speedup of $35\%$.

In summary, we proved that AI assistance in tracking is a good approach, and leads to improvements in tracking code speed and efficiency. 
Using AI leads to a very small and simple codebase, comprised of composing track candidates and feeding them to the neural networks. 
We also found that performance keeps improving with constant training on new data. We intend to continue this development by extending 
the approach to other tracking detectors of the CLAS12, and possibly try to adapt our approach for other experimental detector setups at Jefferson Lab.

\section{Acknowledgments}

This material is based upon work supported by the U.S. Department of Energy, Office of Science, Office of Nuclear 
Physics under contract DE-AC05-06OR23177, and NSF grant no. CCF-1439079 and the Richard T. Cheng Endowment. 
This work was performed using the Turing and  Wahab computing clusters at Old Dominion University.
 
\newpage
\bibliography{references}
\bibliographystyle{ieeetr}

\end{document}